\documentstyle[preprint,aps,epsf]{revtex}
\tighten
\begin{document}
\draft
\title{Scalar Curvature Fluctuations on the Four--Sphere}
\author{S\'{e}rgio M. C. V. Gon\c{c}alves and Ian G. Moss}
\address{
Department of Physics, University of Newcastle Upon Tyne, NE1 7RU U.K.
}
\date{February 1997}
\maketitle
\begin{abstract}
Two-point functions of the scalar curvature for metric fluctuations on
the four--sphere are analysed. The two-point function for points
separated by a fixed distance and for metrics of fixed volume is
calculated using spacetime foam methods. This result can be used for
comparison between the continuum approach to quantum gravity and
numerical quantum gravity on the lattice.
\end{abstract}
\narrowtext
\section{INTRODUCTION}

The aim of this paper is to compute two-point functions for the scalar
curvature   on the 4--sphere. Our results are based upon the de Sitter
space graviton propagator found by Allen and Turyn\cite{allen}.
Curvature two-point functions are also well--suited to evaluation by
numerical quantum gravity on the lattice, therefore the result should
be useful for comparing two different approaches to quantum
gravity\cite{hamber,smit}.

We will use the path integral approach to quantum gravity and constrain
the path integral to Euclidean metrics with a fixed volume. This is
basically the same setup as the theory of spacetime foam\cite{hawking}.
The 4--sphere arises naturally as an instanton in the path integral.

We find that the two--point function $\langle R(x)R(x')\rangle$ is a
constant plus a $\delta-$function term, which generalizes the known
result for flat background\cite{mod,mod2}. This has a simple
explanation because, in quantum field theory, we expect expectation
values values of field equations to vanish. The field equations for
Einstein gravity with a cosmological constant imply that $R=4\Lambda$,
therefore we expect that $\langle R(x)R(x')\rangle= 16\Lambda^2$ for
$x\ne x'$.

In a coordinate independent approach to quantum gravity, the usual
two-point function looses its meaning and it is preferable to examine
the two--point function which has a fixed geodesic distance between the
two points. This propagator has a more complicated functional
dependence which we calculate below. The analytic result shares the
qualitative feature of results from numerical calculations.

\section{THE SCALAR CURVATURE TWO-POINT FUNCTION}

In the Euclidean approach to quantum gravity\cite{hawking}, we examine
the quantum fluctuations of the metric ${\bf \hat{g}}$ about a
classical background ${\bf g}$ of signature $(++++)$,
\begin{equation}
{\bf \hat{g}} = {\bf g}+2\kappa\mbox{\boldmath$\gamma$}
\end{equation}
where $\kappa^2=8\pi G$. The background will be a 4--sphere of radius
$\rho$, related to the cosmological constant by $\rho^2=3/\Lambda$.
Vacuum expectation values are defined by Euclidean path integrals.

The graviton propagator ${\bf G}$ is defined by
\begin{equation}
G^{abc'd'}(x,x')=\langle 0|\gamma^{ab}(x)\gamma^{c'd'}(x')|0\rangle
\end{equation}
It can be decomposed into four components according to its behaviour
under traces and divergences\cite{allen}
\begin{equation}
G^{aba'b'}=G_{TT}{}^{aba'b'}+G_T{}^{aba'b'}+G_L{}^{aba'b'}+G_{PT}{}^{aba'b'}.
\end{equation}
We will use only the following properties of the various components for
contractions with background metric and covariant derivative
\begin{eqnarray}
G_{PT}{}^{aba'b'}&=&-\case1/4g^{ab}g^{a'b'}G\label{gpt}\\
g_{ab}G_{TT}{}^{aba'b'}&=&g_{ab}G_T{}^{aba'b'}=g_{ab}G_L{}^{aba'b'}=0\\
\nabla_a\nabla_bG_L{}^{aba'b'}&=&\left(\nabla^{a'}\nabla^{b'}
-\case1/4g^{a'b'}\nabla^2\right)G\\
\nabla_aG_{TT}{}^{aba'b'}&=&
\nabla_a\nabla_bG_T{}^{aba'b'}=0\label{gradg}
\end{eqnarray}
Here, $G(x,x')$ is a scalar propagator on the 4--sphere which satisfies
\begin{equation}
(-\nabla^2-2\Lambda)G(x,x')=\delta(x,x')\label{speq}
\end{equation}
It depends only on the geodesic separation $\sigma$ of $x$ and $x'$ and
is given explicitly by
\begin{equation}
G(x,x')=-{1\over 2\pi\rho^2}\,{\rm sec}(\pi\nu)\,
F(\case3/2+\nu,\case3/2-\nu,2;z)
\end{equation}
where $\nu^2=33/4$ and
\begin{equation}
z=\cos^2{\sigma\over 2\rho}.
\end{equation}

The first order terms in expansion of the scalar curvature are
\begin{equation}
R(\hat g)=R(g)+2\kappa^2\Delta_{ab}\gamma^{ab}\label{rexp}
\end{equation}
where
\begin{equation}
\Delta_{ab}=\nabla_a\nabla_b-\nabla^2g_{ab}
-\gamma^{ab}R_{ab}(g)
\end{equation}
Therefore the connected two--point function is
\begin{equation}
\langle R(x)R(x')\rangle=16\Lambda^2
+4\kappa^2\Delta_{ab}\Delta_{a'b'}G^{aba'b'}
\end{equation}
From equations (\ref{gpt}-\ref{speq}),
\begin{equation}
\langle R(x)R(x')\rangle=16\Lambda^2
+\case3/2\nabla^2\delta(x,x')+2\Lambda\delta(x,x')\label{rr}
\end{equation}
This result reduces at $\Lambda=0$ to the result for flat space derived
by Modanese \cite{mod,mod2}.

\section{THE TWO-POINT FUNCTION OVER A FIXED DISTANCE}

The two--point function over a fixed geodesic proper distance requires
a coordinate shift in one of the points to compensate for the metric
change in the geodesic length. We do this by moving the second point
along the original geodesic.

Consider, for the background metric $g_{ab}$, a geodesic segment
$x(t)$, from $x$ to $x'$ with proper length given by
\begin{equation}
\sigma_{g}(x,x') = \int_0^1\sqrt{g_{ab}\dot{x}^{a}\dot{x}^{b}} dt.
\end{equation}
The tangent vector $\sigma^{a'}$ is related to $\sigma_{g}(x,x')$ by
\begin{equation}
\sigma^{a'}=g^{a'b'}{\partial\sigma_g\over\partial x^{b'}}
\label{tang}
\end{equation}

Now, for the perturbed metric $\hat g_{ab}$ the geodesic length changes
to
\begin{equation}
\sigma_{\hat{g}}(x,x') = \sigma_{g}(x,x') +\kappa\int_{0}^{\sigma_g}
\gamma_{a''b''}\sigma^{a''}\sigma^{b''}ds
\end{equation}
where $s$ is the proper length along the geodesic to an intermediate
point $x''$. Hence, one has to introduce an appropriate displacement
$\delta x^{a'}$, such that
\begin{equation}
\sigma_{\hat{g}}(x^{a},x^{a'}+\delta
x^{a'})=\sigma_{g}(x^{a},x^{a'})=\sigma.
\end{equation}
Since we also want to remain in the direction of the original geodesic,
we take
\begin{equation}
\delta x^{a'}=\delta x'\sigma^{a'}.
\end{equation}
Combining these equations and using (\ref{tang}) gives
\begin{equation}
\delta x' =
-\kappa\int_{0}^{\sigma} \gamma^{a''b''}
\sigma_{a''}\sigma_{b''}\,ds\label{dx}
\end{equation}
where the integral extends from one end of the geodesic to the other.

The difference between the two-point functions will be denoted by
\begin{equation}
c(x,x')=\langle 0|R(x)R(x'+\delta x')|0\rangle -\langle
0|R(x)R(x')|0\rangle
\end{equation}
To first-order in $\delta x'$,
\begin{equation}
c(x,x')=\langle 0|R(x)\delta x'\sigma^{a'}\nabla_{a'}R(x')|0\rangle.
\end{equation}
This can be computed to order $\kappa^2$ using equations (\ref{rexp})
and (\ref{dx}),
\begin{equation}
c(x,x')= -8\kappa^2\Lambda\int_{0}^{\sigma} ds\,
\langle 0|\gamma^{a''b''}\sigma_{a''}\sigma_{b''}\sigma^{c'}\nabla_{c'}
\Delta_{a'b'}\gamma^{a'b'}|0\rangle
\end{equation}
This simplifies to
\begin{equation}
c(x,x')=-\int_{0}^{\sigma} ds \,
{d\over d\sigma}F(\sigma-s)
\end{equation}
where
\begin{equation}
F(\sigma) =
8\kappa^2\Lambda\,\sigma_a\sigma_b\Delta_{a'b'}G^{aba'b'}(x,x').
\end{equation}
The value of $F(\sigma)$ diverges as $\sigma\to0$ and the integral has
to be regularised. Performing the integral gives
\begin{equation}
c(x,x')=F(\sigma)-F_0
\end{equation}
where $F_0$ is a constant that depends on the regularisation
proceedure. We will absorb $F_0$ as an order $\kappa^2$ correction to
$\Lambda$.

For the evaluation of $F(\sigma)$, we use an identity from Allen and
Turyn\cite{allen},
\begin{equation}
(\nabla^{a'}\nabla^{b'}-\case1/4g^{a'b'}\nabla^2)G=\rho^{-2}
(\sigma^{a'}\sigma^{b'}-\case1/4g^{a'b'})\,z(1-z){d^2G\over dz^2}
\end{equation}
With equations (\ref{gpt}-\ref{gradg}), we obtain
\begin{equation}
c(x,x')=
-16\kappa^2\Lambda^2G(z)-4\kappa^2\Lambda^2(1-2z){dG\over dz}
\end{equation}
This has been plotted in figure \ref{fig2}.

\section{SIMPLICIAL QUANTUM GRAVITY}

The analytic result for the two--point function can be compared with
numerical results obtained from the method of dynamical
triangulations\cite{smit}, a lattice approach to quantum gravity. In
this approach, essentially a modified version of Regge
calculus\cite{regge}, a smooth 4-manifold is approximated by a
simplicial manifold built from equilateral 4-simplices with identical
edge lengths\cite{migdal,weingarten}. In addition, the space inside the
4-simplices is flat, with the curvature being concentrated on
2-simplices.

The path integral over metrics on a manifold with spacetime volume $V$
is defined by a  partition function
\begin{equation}
Z(N_4,k_2) = \sum e^{k_{2}N_{2}}
\end{equation}
where the sum extends over all ways of glueing $N_4$ 4-simplices
together, such that the resulting complex has the topology of a
4-sphere and $N_2$ is the number of 2-simplices in the triangulation.
The parameters relate to the physical volume $V$ and edge length $a$
through
\begin{eqnarray}
k_2&=&\sqrt{3\pi^2\over4}\left({a\over\kappa}\right)^2\label{kd}\\
N_4&=&{96\over\sqrt{5}}\left({V\over a^4}\right)\label{nd}
\end{eqnarray}
The numerical results show different phases, but we concentrate here on
the one which has been interpreted as Einstein gravity \cite{smit}.

In the Euclidean path integral approach to quantum gravity, the
partition function at fixed volume is defined by a Laplace transform in
terms of the partition function with a cosmological
constant,\cite{hawking},
\begin{eqnarray}
Z(V)&=&\int_{\cal C}d\Lambda\,Z(\Lambda)\,\exp(\kappa^{-2}\Lambda V)\\
Z(\Lambda)&=&\int d\mu[g]\,\exp(-I[g]-\kappa^{-2}\Lambda V[g])
\end{eqnarray}
where $d\mu[g]$ is the measure, $I[g]$ is the Einstein--Hilbert action
and $V[g]$ is the volume of the metric. We will have to make
assumptions about the existance of a suitable integral contour ${\cal
C}$.

To leading order in Planck's constant, $Z(\Lambda)$ is dominated by the
4-sphere. This enables us to express any expectation value
$\langle0|f(\gamma)|0\rangle_V$ obtained from the path integral with
constant volume as
\begin{equation}
\langle0|f(\gamma)|0\rangle_V={1\over Z(V)}\int_{\cal C}d\Lambda
\,\langle0|f(\gamma)|0\rangle\,\exp(24\pi^2\kappa^{-2}\Lambda^{-1}
+\kappa^{-2}V\Lambda)
\end{equation}
This integral has a saddle point $\Lambda=\bar\Lambda$, where
$\bar\Lambda^2=24\pi^2/V$. The steepest descents contour is a circle
centered on the origin. The saddle point gives a good approximation for
$V\gg\kappa^4$ and in this case any two-point function calculated at
constant volume can be replaced by the two-point function on the
corresponding 4-sphere.

Numerical results show a two-point function that begins positive for
small distances, becomes negative and then returns positive. The
analytic result also has these features. However, from equations
(\ref{kd}-\ref{nd}),
\begin{eqnarray}
\kappa^2\bar\Lambda&\approx&274.4\,k_2^{-1}N_4^{-1/2}\label{le}\\
\bar\rho/a&\approx&0.1725\,N_4^{1/4}.
\end{eqnarray}
The distance scale in figure \ref{fig2} converted into lattice units
differs by a factor $\approx4$ from the numerical results. It could be
that our identification of the bare parameters in the two approaches
has been too naive, or that the numerical results are still to far
removed from the perturbative regime. It would help resolve this issue
if the numerical results were extended to significantly larger numbers
of simplices. It would also be interesting to compare the analytic
approximation with numerical results for other topologies, particularly
$S^2\times S^2$ which is the next simplest case to the 4-sphere.

\acknowledgments

S.G. is supported by the Programa PRAXIS XXI of the J. N. I. C. T. of
Portugal.

\begin{figure}
\begin{center}
\leavevmode
\epsfxsize=22pc
\epsffile{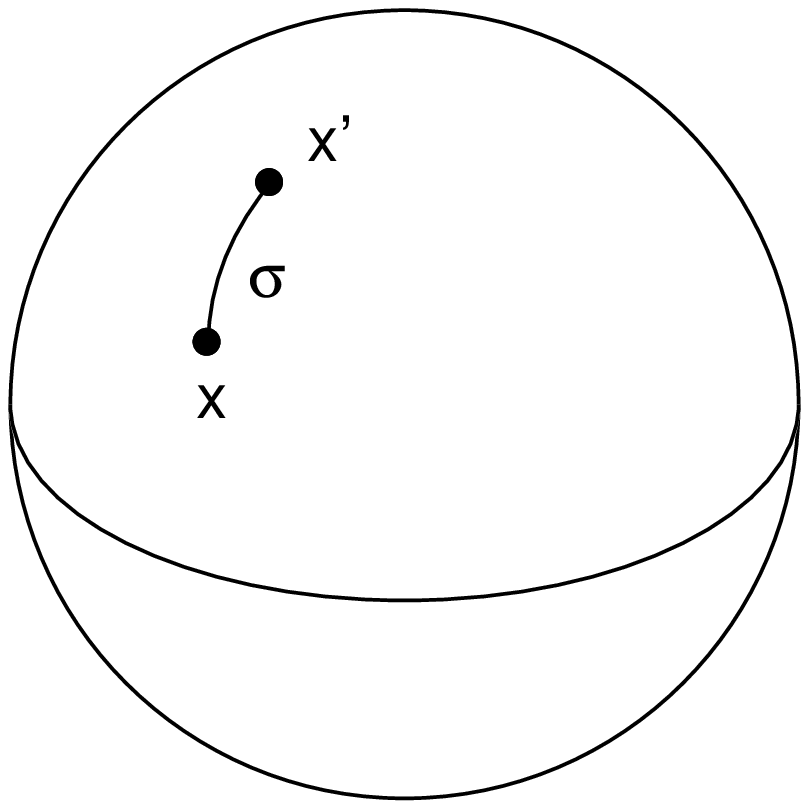}
\end{center}
\caption{A geodesic from $x$ to $x'$ with the tangent vector
$\sigma^a$.\label{fig1}}
\end{figure}

\begin{figure}
\begin{center}
\leavevmode
\epsfxsize=30pc
\epsffile{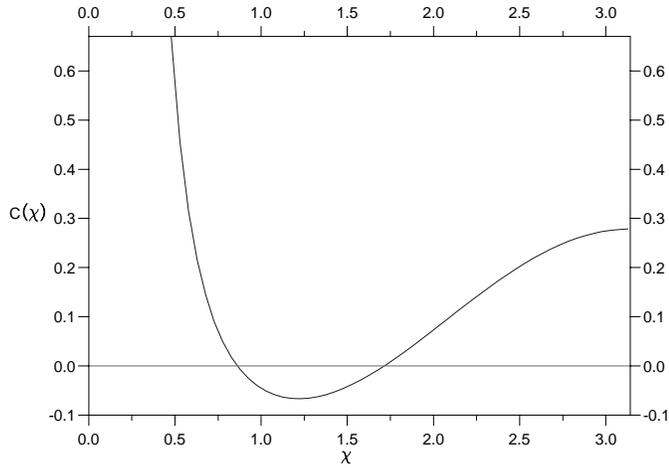}
\end{center}
\caption{A plot of the scalar curvature two-point function on the four
sphere background with the distance $\sigma$ fixed.
$C(\chi)=(\rho/\kappa)^2(16\Lambda^2)^{-2}c(x,x')$ and
$\chi=\sigma/\rho$, where $\rho$ is the radius of the 4--sphere.
\label{fig2}}
\end{figure}


\begin{references}
\bibitem{allen} Allen, B. and Turyn, M. 1987 {\it Nucl. Phys. B} {\bf
292} 813

\bibitem{hamber}Hamber H. W. 1992 {\it Phys. Rev.} {\bf D45} 507; 1993
{\it Nucl. Phys.} {\bf B400} 347; 1995 {\it Phys. rev.} {\bf D50}
3942.

\bibitem{smit} Smit, J. and Debakker, B. V. 1995 {\it Nucl. Phys. B}
{\bf 454} 343

\bibitem{hawking} Hawking, S. W. 1979 in {\it General Relativity: An
Einstein Centenary Survey} Eds Hawking, S. W. and Israel, W. Cambridge
University Press: Cambridge

\bibitem{mod} Modanese, G. 1992 {\it Phys. Lett. B} {\bf 288} 69

\bibitem{mod2} Modanese, G. 1994 {\it Riv. Nuovo Cimento} {\bf 17} No.8
1

\bibitem{regge} Regge, T. 1961 {\it Il Nuovo Cimento} {\bf XIX} No. 3
558

\bibitem{migdal}Agishtein M. E. and Migdal A. A. 1992 {\it Mod. Phys.
Lett.} {\bf A7} 1039
\bibitem{weingarten}Weingarten D. 1982 {\it Nucl. Phys.} {\bf B210} 229


\end{references}
\end{document}